\documentclass[aps,prd,preprint,groupedaddress,preprintnumbers]{revtex4-1}

\newcommand{\e}{\mathrm{e}}
\newcommand{\ep}{\epsilon}
\newcommand{\vev}[1]{\left\langle #1 \right\rangle}
\newcommand{\vvev}[1]{\left\langle\kern-0.3em\left\langle #1
      \right\rangle\kern-0.3em\right\rangle} 
\newcommand{\Op}{\mathcal{O}}
\newcommand{\lb}{\left\lbrace}
\newcommand{\rb}{\right\rbrace}
\newcommand{\N}{\mathcal{N}}
\newcommand{\nn}{\nonumber}

\begin{document}

\preprint{KOBE-TH-15-03}

\title{Construction of the Energy-Momentum Tensor \\for Wilson Actions}


\author{H. Sonoda}
\email[]{hsonoda@kobe-u.ac.jp}
\affiliation{Physics Department, Kobe University, Kobe 657-8501, Japan}


\date{11 April 2015, revised 1 September 2015}

\begin{abstract}
    Given an arbitrary Wilson action of a real scalar field, we
    discuss how to construct the energy-momentum tensor of the theory.
    Using the exact renormalization group, we can determine the
    energy-momentum tensor implicitly, but we are short of obtaining
    an explicit formula in terms of the Wilson action.
\end{abstract}


\maketitle

\section{Introduction\label{section-introduction}}

The energy-momentum tensor is important for quantum field theory; it
describes the conservation of energy, momentum, and angular momentum,
and it also determines how the theory couples to weak external
gravity.  Long ago the construction of a symmetric energy-momentum
tensor was discussed extensively in the canonical formalism by
Belinfante \cite{Belinfante} and Rosenfeld \cite{Rosenfeld}.  It was
shown that, given a lagrangian, an explicit formula can be given for
the energy-momentum tensor.  More recently (but also long ago) the
construction of the ``improved'' energy-momentum tensor was discussed
in \cite{Callan:1970ze}, where renormalizability of the
energy-momentum tensor in renormalized theories was proven within
perturbation theory.

In this paper we are interested in constructing the energy-momentum
tensor of a generic real scalar theory, given its Wilson action.  A
Wilson action incorporates a momentum cutoff so that its functional
integrals are well defined.  But it contains all possible local terms,
and no canonical method can be adopted for the construction of the
energy-momentum tensor.  The most straightforward approach would be to
couple the theory to external gravity, and define the energy-momentum
tensor as the functional derivative of the action with respect to the
external metric.  However, this adds further complexity to the study,
and we would like to remain in flat space if possible.

In constructing the energy-momentum tensor, we need an approach
suitable to Wilson actions.  Local composite operators, introduced in
\cite{Wilson:1973jj} (sect.~12.4 and Appendix) and
\cite{Becchi:1996an}, are infinitesimal deformations of a Wilson
action, and the energy-momentum tensor is a particular example.  We
will formulate the Ward identity for the energy-momentum tensor using
composite operators, in particular, equation-of-motion composite
operators.\cite{Igarashi:2009tj} We will see that the
equation-of-motion composite operators give us sufficient information
to determine the symmetric energy-momentum tensor.  We are short of
obtaining an explicit formula in terms of the action, but we are left
with only the familiar ambiguity that would correspond to the coupling
to the Riemann curvature tensor if external gravity were
present.\cite{Polchinski:1987dy}

Throughout the paper, we resort to the assumption of locality of
composite operators: if the space integral of a local composite
operator vanishes, the operator must be a derivative.  In the Fourier
space, if a local composite operator vanishes at zero momentum, it
must be proportional to a momentum:
\begin{equation}
\Op (p=0) = 0 \Longrightarrow \Op (p) = p_\mu J_\mu (p)\label{locality}
\end{equation}
where $J_\mu (p)$ is also a local composite operator.

Our present work was motivated by a preliminary presentation of the
work \cite{Delamotte:2015aaa} in which the conformal invariance
\cite{Wess:1960} of the $O(N)$ invariant scalar theory at its fixed
point was proven using Wilson actions and the exact renormalization
group.  In a recent work by Rosten \cite{Rosten:2014oja} the
energy-momentum tensor has been constructed for scale invariant Wilson
actions.  Our section \ref{section-fp} has some overlap with part of
the results obtained in \cite{Rosten:2014oja}.

This paper is organized as follows.  In sect.~\ref{section-wilson} we
briefly review Wilson actions and the exact renormalization group
(ERG) transformation following the results of
\cite{Sonoda:2015bla}. Additional details are given in Appendices
\ref{appendix-EOM} \& \ref{appendix-ERGdiff}.  In
sect.~\ref{section-inv} we derive the operator equations starting from
the naive invariance of the Wilson action under translations and
rotations.  In sect.~\ref{section-Theta} we derive the Ward identity
for translation invariance that determines a symmetric energy-momentum
tensor implicitly.  In sect.~\ref{section-consistency} we use ERG to
constrain the trace of the energy-momentum tensor.  In
sect.~\ref{section-fp} we discuss the energy-momentum tensor for a
fixed point Wilson action.  Under the assumption, formulated in
\cite{Polchinski:1987dy}, we derive the Ward identity for conformal
invariance. In sect.~\ref{section-gaussian} we show how our
construction works using the massless free theory as an example.  We
conclude the paper in sect.~\ref{section-conclusions}.

Throughout the paper we work in $D$-dimensional Euclidean momentum
space, and we use the abbreviated notation
\begin{equation}
\int_p = \int \frac{d^D p}{(2 \pi)^D},\quad
\delta (p) = (2 \pi)^D \delta^{(D)} (p)
\end{equation}

\section{Wilson Actions\label{section-wilson}}

We consider a real scalar theory in $D$-dimensional Euclidean space.
We denote the Fourier transform of the scalar field by $\phi (p)$.
Let $S[\phi]$ be a Wilson action that incorporates a momentum cutoff
so that the correlation functions defined by
\begin{equation}
\vev{\phi (p_1) \cdots \phi (p_n)}_S \equiv
\int [d\phi]\, \phi (p_1) \cdots \phi (p_n)\, \e^{S [\phi]}
\label{vev}
\end{equation}
are free from ultraviolet divergences.  We measure all dimensionful
quantities in units of an appropriate power of the cutoff and deal
only with dimensionless quantities.  Hence, our momentum cutoff
becomes $1$ in this convention.

We introduce a positive cutoff function $K (p)$ that depends on the squared
momentum $p^2 = p_\mu p_\mu$ and decreases rapidly for $p^2 > 1$.  We
impose
\begin{equation}
K(p) \stackrel{p^2 \to 0}{\longrightarrow} 1
\end{equation}
so that the cutoff does not affect low energy physics.  Using $K (p)$,
we define modified correlation functions by
\begin{eqnarray}
&&\vvev{\phi (p_1) \cdots \phi (p_n)}_S \equiv \prod_{i=1}^n
\frac{1}{K(p_i)}\nn\\
&&\quad \times 
\vev{\exp \left( - \int_p \frac{K(p)\left(1-K(p)\right)}{p^2}
      \frac{1}{2} \frac{\delta^2}{\delta \phi (p) \delta \phi (-p)}
  \right) \phi (p_1) \cdots \phi (p_n)}_S
\label{vvev}
\end{eqnarray}
The exact renormalization group (ERG) transformation $R_t$ acting on
$S$ is defined so that the modified correlation functions are simply
related as
\begin{equation}
\vvev{\phi (p_1 \e^t) \cdots \phi (p_n \e^t)}_{R_t S}
= \e^{t \cdot n \left(- \frac{D+2}{2} + \gamma \right)}
\vvev{\phi (p_1) \cdots \phi (p_n)}_S\label{equivalence}
\end{equation}
where the anomalous dimension $\gamma$ of $\phi$ is chosen so that
$R_t$ has a fixed point.\cite{Sonoda:2015bla}

Besides the Wilson action, local composite operators play a very
important role in this paper.  A local composite operator $\Op (p)$ is
a functional of $\phi$ which can be interpreted as an infinitesimal
deformation of the Wilson action.\cite{Wilson:1973jj,Becchi:1996an} We
define its modified correlation functions by
\begin{eqnarray}
&&\vvev{\Op (p)\,\phi (p_1) \cdots \phi (p_n)}_S
\equiv \prod_{i=1}^n \frac{1}{K (p_i)}\nn\\
&&\quad \times \vev{ \Op (p) 
\exp \left( - \int_q \frac{K(q)\left(1-K(q)\right)}{q^2}
      \frac{1}{2} \frac{\delta^2}{\delta \phi (q) \delta \phi (-q)}
  \right) \phi (p_1) \cdots \phi (p_n)}_S
\end{eqnarray}
ERG acts on $\Op$ such that
\begin{equation}
\vvev{\left(R_t \Op\right) (p \e^t)\,
\phi (p_1 \e^t) \cdots \phi (p_n \e^t)}_{R_t S}
= \e^{t  n \left(-\frac{D+2}{2} + \gamma\right)} \vvev{\Op
  (p)\,\phi (p_1) \cdots \phi (p_n)}_{S}
\label{Rt-composite}
\end{equation}
(See Appendices \ref{appendix-EOM} and \ref{appendix-ERGdiff} for more
details.) 

\section{Invariance under Translations and Rotations\label{section-inv}}

We assume that the theory described by the Wilson action $S[\phi]$ is
invariant under both translations and rotations.  Let us derive the
consequences of this assumption.  Some relevant technical details on
composite operators are given in Appendix \ref{appendix-EOM}.

\subsection{Translation Invariance}

Translation invariance implies that the correlation functions satisfy
the Ward identity
\begin{equation}
\sum_{i=1}^n p_{i\mu} \vvev{\phi (p_1) \cdots \phi (p_n)}_S = 0
\end{equation}
This is equivalent to the operator identity
\begin{equation}
\int_q K(q) \e^{-S} \frac{\delta}{\delta \phi (q)} \left( q_\mu \left[
        \phi (q)\right] \e^S \right) = 0
\label{translation-inv}
\end{equation}
where
\begin{equation}
\left[ \phi (q) \right] \equiv \frac{1}{K(q)} \left( \phi (q) +
    \frac{K(q)\left(1 - K(q)\right)}{q^2} \frac{\delta S}{\delta \phi
      (-q)} \right)
\end{equation}
is the composite operator satisfying
\begin{equation}
\vvev{\left[ \phi (q)\right] \phi (p_1) \cdots \phi (p_n)}_S
= \vvev{\phi (q) \phi (p_1) \cdots \phi (p_n)}_S
\end{equation}

In the following we wish to show, in some details, that the operator
identity (\ref{translation-inv}) is a consequence of the naive
translation invariance
\begin{equation}
\int_q q_\mu \phi (q) \frac{\delta S}{\delta \phi (q)} = 0
\label{bare-translation-inv}
\end{equation}
which can be rewritten as
\begin{equation}
\int d^D r\, \partial_\mu \phi (r) \frac{\delta S}{\delta \phi (r)} =
0
\end{equation}
in coordinate space.  The derivation is straightforward.  Defining
\begin{equation}
k(q) \equiv K(q)\left(1 - K(q)\right)
\end{equation}
for simplicity, we find
\begin{eqnarray}
&&\int_q K(q) \e^{-S} \frac{\delta}{\delta \phi (q)} \left(
q_\mu \left[ \phi (q)\right]\,\e^S \right)\nn\\
&&= \int_q \e^{-S} \frac{\delta}{\delta \phi (q)} \left[
    q_\mu \left( \phi (q) + \frac{k(q)}{q^2}
        \frac{\delta S}{\delta \phi (-q)} \right) \,\e^S \right]\nn\\
&&= \int_q \left[ q_\mu \delta (0) + q_\mu \frac{k(q)}{q^2}
    \frac{\delta^2 S}{\delta \phi (q) \delta \phi (-q)} + q_\mu \left(
        \phi (q) + \frac{k(q)}{q^2} \frac{\delta S}{\delta \phi
          (-q)}\right) \frac{\delta S}{\delta \phi (q)} \right]\nn\\
&&= \int_q q_\mu \phi (q) \frac{\delta S}{\delta \phi (q)} 
 + \int_q q_\mu \frac{k(q)}{q^2} \lb \frac{\delta^2 S}{\delta \phi
  (q) \delta \phi (-q)} + \frac{\delta S}{\delta \phi (-q)}
\frac{\delta S}{\delta \phi (q)} \rb
\end{eqnarray}
The first term vanishes due to (\ref{bare-translation-inv}),
and the second term vanishes since the integrand is odd under the
inversion $q \to - q$.  Hence, we obtain (\ref{translation-inv}).

\subsection{Rotation Invariance}

Rotation invariance implies the Ward identity
\begin{equation}
\ep_{\mu\nu} \sum_{i=1}^n p_{i \nu} \vvev{\phi (p_1) \cdots
  \frac{\partial}{\partial p_{i\mu}} \phi (p_i) \cdots \phi (p_n)}_S =
0
\end{equation}
for an arbitrary constant antisymmetric tensor $\ep_{\mu\nu} = -
\ep_{\nu\mu}$.  This is equivalent to the operator identity
\begin{equation}
\ep_{\mu\nu} \int_q K(q) \e^{- S} \frac{\delta}{\delta \phi (q)}
\left( q_\nu \frac{\partial}{\partial q_\mu} \left[ \phi (q)\right]
    \e^S \right) = 0\label{rotation-inv}
\end{equation}
We wish to show, in the following, that this operator identity results
from the naive rotation invariance of the action:
\begin{equation}
\int_q \ep_{\mu\nu} q_\nu \frac{\partial \phi (q)}{\partial q_\mu}
\frac{\delta S}{\delta \phi (q)} = 0 \label{bare-rotation-inv}
\end{equation}
The derivation is again straightforward.

Using the rotation invariance of $k(q) = K(q)\left(1-K(q)\right)$, we
obtain
\begin{eqnarray}
&&\int_q K(q) \e^{-S} \ep_{\mu\nu} q_\nu \frac{\delta}{\delta \phi (q)}
\left( \frac{\partial}{\partial q_\mu} \left[ \phi (q) \right]\, \e^S
\right)\nn\\
&&\quad = \int_q \e^{-S} \frac{\delta}{\delta \phi (q)}
\left[ \ep_{\mu\nu} q_\nu \frac{\partial}{\partial q_\mu} \left(
        \phi (q) + \frac{k(q)}{q^2} \frac{\delta S}{\delta \phi (-q)}
    \right) \e^S \right]\nn\\
&&\quad = \int_q \frac{k(q)}{q^2} \left(\ep_{\mu\nu} q_\nu
\frac{\partial}{\partial q_\mu} \frac{\delta^2 S}{\delta \phi (-q)
  \delta \phi (q')}\right)_{q' \to q}\nn\\
&&\qquad+ \int_q \ep_{\mu\nu} q_\nu 
    \frac{\partial \phi (q)}{\partial q_\mu} 
\frac{\delta S}{\delta \phi (q)} \e^S
+ \int_q \frac{k(q)}{q^2} \ep_{\mu\nu} q_\nu
    \frac{\partial}{\partial q_\mu} \frac{\delta S}{\delta \phi (-q)}
    \cdot \frac{\delta S}{\delta \phi (q)}
\end{eqnarray}
The second term on the right vanishes due to
(\ref{bare-rotation-inv}).  Hence, we obtain
\begin{eqnarray}
&&\int_q K(q) \e^{-S} \ep_{\mu\nu} q_\nu \frac{\delta}{\delta \phi (q)}
\left( \frac{\partial}{\partial q_\mu} \left[ \phi (q) \right]\, \e^S
\right)\nn\\
&&\quad=  \int_q \frac{k(q)}{q^2} \frac{1}{2} \ep_{\mu\nu} q_\nu
\frac{\partial}{\partial q_\mu} \left( \frac{\delta^2 S}{\delta \phi
      (-q) \delta \phi (q)} + \frac{\delta S}{\delta \phi
      (-q)} \frac{\delta S}{\delta \phi (q)} \right)\nn\\
&&\quad = - \int_q \ep_{\mu\nu} q_\nu
\frac{\partial}{\partial q_\mu} \frac{k(q)}{q^2} \cdot \frac{1}{2}
\left( \frac{\delta^2 S}{\delta \phi 
      (-q) \delta \phi (q)} + \frac{\delta S}{\delta \phi
      (-q)} \frac{\delta S}{\delta \phi (q)} \right) 
\end{eqnarray}
This vanishes thanks to the rotation invariance of $k(q)$.  Hence, we
obtain (\ref{rotation-inv}).

\section{Construction of the Energy-Momentum Tensor\label{section-Theta}}

Given a Wilson action $S[\phi]$ with the invariance under translations
and rotations, we would like to construct the energy-momentum tensor.

To start with, we introduce a local composite operator with momentum
$p$, defined by
\begin{equation}
J_\mu (p) \equiv \int_q K(q) \e^{-S} \frac{\delta}{\delta \phi (q)}
\left( (p+q)_\mu \left[ \phi (p+q)\right] \e^S \right)
\end{equation}
This vanishes at $p=0$ as a consequence of the translation invariance
(\ref{translation-inv}):
\begin{equation}
J_\mu (0) = 0
\end{equation}
Since $J_\mu (p)$ is a local composite operator that vanishes at $p =
0$, it must be proportional to $p$ so that
\begin{equation}
J_\mu (p) = p_\nu \Theta_{\nu\mu} (p) \label{J-EM}
\end{equation}
in terms of another local composite operator $\Theta_{\nu\mu} (p)$,
which we call the energy-momentum tensor.  Note that this relation
does not determine $\Theta_{\nu\mu} (p)$ uniquely; it has an additive
ambiguity by
\begin{equation}
p_\alpha Y_{\nu\alpha,\mu} (p)
\end{equation}
where 
\begin{equation}
Y_{\alpha\nu,\mu} = - Y_{\nu\alpha,\mu}
\end{equation}

We now differentiate (\ref{J-EM}) with respect to $p_\alpha$ to obtain
\begin{equation}
\Theta_{\alpha\mu} (p) + p_\nu \frac{\partial}{\partial p_\alpha}
\Theta_{\nu\mu} (p) = \int_q K(q) \e^{-S} \frac{\delta}{\delta \phi
  (q)} \left[ \lb \delta_{\alpha\mu} + (p+q)_\mu
    \frac{\partial}{\partial q_\alpha} \rb \left[ \phi (p+q)\right]
    \e^S \right]
\end{equation}
Antisymmetrizing this with respect to $\alpha$ and $\mu$, we obtain
\begin{eqnarray}
&&\Theta_{\alpha\mu} (p) - \Theta_{\mu\alpha} (p) + p_\nu \left(
    \frac{\partial \Theta_{\nu\mu} (p)}{\partial p_\alpha} -
    \frac{\partial \Theta_{\nu\alpha} (p)}{\partial p_\mu}
\right)\nn\\
&&= \int_q K(q) \e^{-S} \frac{\delta}{\delta \phi (q)} \left[
\lb (q+p)_\mu \frac{\partial}{\partial q_\alpha} - (q+p)_\alpha
\frac{\partial}{\partial q_\mu} \rb \left[ \phi (p+q)\right] \e^S
\right] 
\end{eqnarray}
Rotation invariance implies the vanishing of the right-hand side in
the limit $p = 0$.  Hence, we find
\begin{equation}
\Theta_{\alpha\mu} (p) - \Theta_{\mu\alpha} (p) \stackrel{p \to
  0}{\longrightarrow} 0
\end{equation}
Since $\Theta_{\mu\nu} (p)$ is a local composite operator, this
implies that the antisymmetric part of $\Theta_{\mu\nu} (p)$ is a
local composite operator proportional to $p$:
\begin{equation}
\Theta_{\mu\nu} (p) - \Theta_{\nu\mu} (p) = p_\alpha \tau_{\alpha,
  \mu\nu} (p)
\end{equation}
where
\begin{equation}
\tau_{\alpha,\mu\nu} (p) = - \tau_{\alpha,\nu\mu} (p)
\end{equation}

Using $\tau_{\alpha,\mu\nu}$ we can construct a symmetric
energy-momentum tensor following the procedure given in
\cite{Belinfante} and \cite{Rosenfeld}. We define
\begin{equation}
B_{\alpha\mu\nu} (p) \equiv \frac{1}{2} \left( \tau_{\alpha,\mu\nu}
    (p) + \tau_{\mu,\nu\alpha} (p) - \tau_{\nu,\alpha\mu} (p) \right)
\end{equation}
which is antisymmetric with respect to the first two indices
\begin{equation}
B_{\alpha\mu\nu} (p) = - B_{\mu\alpha\nu} (p)\label{B-antisym}
\end{equation}
and
\begin{equation}
B_{\alpha\mu\nu} (p) - B_{\alpha\nu\mu} (p) = \tau_{\alpha,\mu\nu} (p)
\label{B-tau}
\end{equation}
We then redefine the energy-momentum tensor by
\begin{equation}
\Theta'_{\mu\nu} (p) \equiv \Theta_{\mu\nu} (p) - p_\alpha
B_{\alpha\mu\nu} (p)
\end{equation}
which, on account of (\ref{B-antisym}), satisfies
\begin{equation}
p_\mu \Theta_{\mu\nu} (p) = p_\mu \Theta'_{\mu\nu} (p)
\end{equation}
and is, thanks to (\ref{B-tau}), symmetric
\begin{equation}
\Theta'_{\mu\nu} (p) = \Theta'_{\nu\mu} (p)
\end{equation}
We will omit the prime on $\Theta'_{\mu\nu}$ from now on.

To summarize so far, our energy-momentum tensor $\Theta_{\mu\nu} (p)$
is symmetric, and it satisfies
\begin{equation}
p_\mu \Theta_{\mu\nu} (p) = \int_q K(q) \e^{-S} \frac{\delta}{\delta
  \phi (q)} \left( (p+q)_\nu \left[\phi (p+q)\right] \e^S \right)
\label{EM-translation}
\end{equation}
for translation invariance, and
\begin{eqnarray}
&& p_\nu \left(
    \frac{\partial \Theta_{\nu\mu} (p)}{\partial p_\alpha} -
    \frac{\partial \Theta_{\nu\alpha} (p)}{\partial p_\mu}
\right)\nn\\
&&= \int_q K(q) \e^{-S} \frac{\delta}{\delta \phi (q)} \left(
\lb (q+p)_\mu \frac{\partial}{\partial q_\alpha} - (q+p)_\alpha
\frac{\partial}{\partial q_\mu} \rb \left[ \phi (p+q)\right] \e^S
\right)\label{EM-rotation}
\end{eqnarray}
for rotation invariance.  (\ref{EM-rotation}) is a consequence of the
symmetry $\Theta_{\mu\nu} = \Theta_{\nu\mu}$ and
(\ref{EM-translation}).  The energy-momentum tensor $\Theta_{\mu\nu}
(p)$ is now left with an additive ambiguity by a symmetric tensor
\begin{equation}
p_\alpha p_\beta Y_{\mu\alpha,\nu\beta} (p)\label{EM-ambiguity}
\end{equation}
where $Y_{\mu\alpha,\nu\beta} (p)$ must satisfy
\begin{equation}
p_\mu p_\alpha p_\beta Y_{\mu\alpha,\nu\beta}(p) = 0
\end{equation}
This implies \cite{Polchinski:1987dy}
\begin{equation}
Y_{\mu\alpha,\nu\beta} = Y_{\nu\beta,\mu\alpha} = -
Y_{\alpha\mu,\nu\beta} = - Y_{\mu\alpha,\beta\nu}
\end{equation}
which is the symmetry of the Riemann curvature tensor $R_{\mu\alpha\nu\beta}$.

\section{Consistency with ERG\label{section-consistency}}

We now combine the results of the previous section with ERG.
Differentiating (\ref{EM-translation}) with respect to $p_\nu$ and
summing over $\nu$, we obtain
\begin{eqnarray}
&&\Theta (p) + p_\mu \frac{\partial}{\partial p_\nu} \Theta_{\mu\nu}
(p)\nn\\
&&= - D \N (p) + \int_q K(q) \e^{-S} \frac{\delta}{\delta \phi (q)}
\left[ (q+p)_\nu \frac{\partial}{\partial q_\nu} \left[ \phi
        (p+q)\right] \e^S \right]
\label{trace}
\end{eqnarray}
where $\Theta \equiv \Theta_{\mu\mu}$ is the trace, and
\begin{equation}
\N (p) \equiv - \int_q K(q) \e^{-S} \frac{\delta}{\delta \phi (q)}
\left( \left[\phi (q+p) \right] \e^S \right)
\end{equation}
is an equation-of-motion composite operator satisfying
\begin{equation}
\vvev{\N (p) \phi (p_1)\cdots \phi (p_n)}_S
= \sum_{i=1}^n \vvev{\phi (p_1) \cdots \phi (p_i+p) \cdots \phi
  (p_n)}_S
\end{equation}
In the limit $p \to 0$, (\ref{trace}) gives
\begin{equation}
\Theta (0) = - D \N + \int_q K(q) \e^{-S} \frac{\delta}{\delta \phi
  (q)} \left[ q_\mu \frac{\partial}{\partial q_\mu} \left[ \phi
        (q)\right] \e^S \right]
\label{Theta-at-zero}
\end{equation}
where we denote $\N \equiv \N (p=0)$. 

The trace $\Theta (0)$ is related to the ERG differential equation,
which is briefly reviewed in Appendix \ref{appendix-ERGdiff}.  ERG
acts on the Wilson action $S$ as
\begin{equation}
\mathcal{D} S [\phi]
\equiv \frac{d}{dt} \left(R_t S\right)\Big|_{t=0}
=  \int_q K(q) \e^{-S [\phi]} \frac{\delta}{\delta
  \phi (q)} \left[ q_\mu \frac{\partial}{\partial q_\mu} \left[ \phi
        (q)\right] \e^{S [\phi]} \right] + \left( - \frac{D+2}{2} + \gamma
\right) \N \label{DS}
\end{equation}
(This is (\ref{ERGdiff-rewritten}) of Appendix \ref{appendix-ERGdiff}.)
Comparing this with (\ref{Theta-at-zero}), we obtain
\begin{equation}
\Theta (0) = \mathcal{D} S - \left( \frac{D-2}{2} + \gamma
\right) \N 
\end{equation}
where $\frac{D-2}{2} + \gamma$ is the full scale dimension of the
scalar field in coordinate space.  Generalizing this to non-vanishing
momenta, we obtain
\begin{equation}
\Theta (p) = \Op (p) - \left( \frac{D-2}{2} + \gamma
\right) \N (p) \label{Theta-at-nonzero}
\end{equation}
where $\Op (p)$ is a local composite operator whose zero momentum
limit is
\begin{equation}
\Op (0) = \mathcal{D} S
\end{equation}
defined by (\ref{DS}).  Please recall that $\Theta_{\mu\nu} (p)$ is
ambiguous by $p_\alpha p_\beta Y_{\mu\alpha,\nu\beta} (p)$.  This
translates into the ambiguity of $\Theta (p)$ (and also $\Op (p)$) by
$p_\alpha p_\beta Y_{\mu\alpha,\mu\beta} (p)$.

\section{The Energy-Momentum Tensor at a Fixed Point\label{section-fp}}

Let us suppose $S$ is a fixed point Wilson action $S^*$ corresponding
to the anomalous dimension $\gamma$ for the scalar field.  We then
find
\begin{equation}
\Op (0) = 0
\end{equation}
so that there must be a local composite operator $K_\mu (p)$ that gives
\begin{equation}
\Op (p) = p_\mu K_\mu (p)
\end{equation}
Hence, (\ref{Theta-at-nonzero}) becomes
\begin{equation}
\Theta (p) = p_\mu K_\mu (p) - \left(\frac{D-2}{2} + \gamma\right) \N
(p) \label{trace-p}
\end{equation}
Especially, at $p=0$, we obtain
\begin{equation}
\Theta (0) = -   \left(\frac{D-2}{2} + \gamma\right) \N
\end{equation}
which has been obtained also in \cite{Rosten:2014oja} (see (4.44) at
the end of IV D of \cite{Rosten:2014oja}).

We now consider local composite operators at the fixed point, which
form an infinite dimensional linear space.  A local composite operator
$\Op$ with scale dimension $-y$ satisfies
\begin{equation}
R_t \Op = \e^{+ y t} \Op\label{diagonalized}
\end{equation}
so that (\ref{Rt-composite}) gives
\begin{equation}
\vvev{\Op (p \e^t)\,\phi (p_1 \e^t) \cdots \phi (p_n \e^t)}_{S^*}
= \e^{t \lb - y + n \left( - \frac{D+2}{2} + \gamma \right) \rb}
\vvev{\Op (p)\,\phi (p_1) \cdots \phi (p_n)}_{S^*}
\end{equation}
at the fixed point.  We assume that we can choose a basis consisting
of composite operators with definite scale dimensions.  An arbitrary
composite operator can then be given as a linear combination of
various composite operators with various scale dimensions.

The translation invariance (\ref{EM-translation}) implies that
$\Theta_{\mu\nu} (p)$ has scale dimension $0$.  (In coordinate space
$\Theta_{\mu\nu} (x) \equiv \int_p \e^{i p x} \Theta_{\mu\nu} (p)$ has
scale dimension $D$.)  Since $\mathcal{N} (p)$ also has scale
dimension $0$, the current $K_\mu (p)$ must have scale dimension $-1$.
(Hence, scale dimension $D-1$ in coordinate space.)  If there exists
no well-defined local composite operator $K_\mu (p)$ that has scale
dimension $-1$, i.e.,
\begin{equation}
\left(R_t K_\mu\right) (p) = \e^{t} K_\mu (p)
\label{Kmu-scale-dim}
\end{equation}
then $\Op (p)$ inevitably vanishes, as has been also pointed out in
the note added of \cite{Rosten:2014oja}.  (The non-trivial
differential equation corresponding to (\ref{Kmu-scale-dim}) is
obtained from (\ref{comp-scale-dim}) of Appendix
\ref{appendix-ERGdiff} with $y=1$.)  For the real scalar theory in
$D=3$, it has recently been shown in \cite{Delamotte:2015aaa} that a
well defined local composite operator $K_\mu (p)$ of scale dimension
$-1$ does not exist so that $\Op (p)$ vanishes.  Hence, as is
explained below, scale invariance implies conformal invariance for the
critical Ising model.

Now, it has been of much interest lately whether scale invariance
implies conformal invariance or not.  (See
\cite{Delamotte:2015aaa,Rosten:2014oja} within the context of ERG.  See also
\cite{Nakayama:2013is} for a recent review.)  As has been shown in
\cite{Polchinski:1987dy}, conformal invariance is equivalent to
\begin{equation}
\Op (p) = p_\mu p_\nu L_{\mu\nu} (p) \label{O-quadratic}
\end{equation}
where $L_{\mu\nu} (p)$ is a symmetric local composite operator of
scale dimension $-2$.  Note that the ambiguity of $\Op (p)$ by
$p_\alpha p_\beta Y_{\mu\alpha,\nu\beta} (p)$ affects $L_{\mu\nu} (p)$
by $Y_{\alpha\mu,\alpha\nu} (p)$.  Hence, it does not affect the
conclusion below.  (In fact as has been shown in
\cite{Polchinski:1987dy} we can redefine $\Theta_{\mu\nu} (p)$ to make
$\Op (p) = 0$ if (\ref{O-quadratic}) holds.)

If (\ref{O-quadratic}) holds, we can derive the Ward identity
for conformal invariance.  In coordinate space the infinitesimal
conformal transformation is given by \cite{Wess:1960}
\begin{equation}
\delta \phi (x) = \ep_\alpha \left( - x_\alpha x_\nu \partial_\nu +
  \frac{1}{2} x^2 \partial_\alpha - \left(\frac{D-2}{2}+\gamma\right)
  x_\alpha \right) \phi (x)
\end{equation}
where the last term proportional to $x_\alpha$ is determined by the
full scale dimension of the scalar field.  Going to the momentum
space, we obtain
\begin{equation}
\delta \phi (p) = i \ep_\alpha \left( p_\nu \frac{\partial^2}{\partial
    p_\nu \partial p_\alpha} - \frac{1}{2} p_\alpha
  \frac{\partial^2}{\partial p_\nu \partial p_\nu} 
+ \left(D - \left(\frac{D-2}{2} + \gamma\right)\right) \frac{\partial}{\partial
  p_\alpha} \right) \phi (p) 
\end{equation}
To derive the Ward identity, we use
\begin{equation}
\Theta (p) = p_\mu p_\nu L_{\mu\nu} (p) - \left(\frac{D-2}{2} + \gamma
\right) \N (p)
\label{Theta-conformal}
\end{equation}
and compute
\begin{eqnarray}
&&p_\mu \left( \frac{\partial^2}{\partial p_\alpha \partial p_\nu} -
   \frac{1}{2} \delta_{\alpha\nu} \frac{\partial^2}{\partial p_\beta \partial
      p_\beta} \right) \Theta_{\nu\mu} (p)\nn\\
&=& \left( \frac{\partial^2}{\partial p_\alpha \partial p_\nu} - \frac{1}{2}
    \delta_{\alpha\nu} \frac{\partial^2}{\partial p_\beta \partial
      p_\beta} \right) \left(p_\mu \Theta_{\nu\mu} (p) \right) - 
\frac{\partial}{\partial p_\alpha} \Theta (p)\nn\\
&=& \left( \frac{\partial^2}{\partial p_\alpha \partial p_\nu} - \frac{1}{2}
    \delta_{\alpha\nu} \frac{\partial^2}{\partial p_\beta \partial
      p_\beta} \right) \int_q K(q) \e^{-{S^*}} \frac{\delta}{\delta \phi
  (q)} \left[ (q+p)_\nu \left[ \phi (q+p)\right] \e^{{S^*}} \right]\nn\\
&&\quad + \frac{\partial}{\partial p_\alpha} \left(
 \left(\frac{D-2}{2} +\gamma\right) \N (p) -  p_\mu p_\nu L_{\mu\nu} (p)\right)
\end{eqnarray}
where we have used (\ref{EM-translation}) and
(\ref{Theta-conformal}).  We then obtain
\begin{eqnarray}
&&p_\mu \left(  \frac{\partial^2}{\partial p_\alpha \partial p_\nu} - \frac{1}{2}
    \delta_{\alpha\nu} \frac{\partial^2}{\partial p_\beta \partial
      p_\beta} \right) \Theta_{\nu\mu} (p)\nn\\
&=& \int_q K(q) \e^{-{S^*}} \frac{\delta}{\delta \phi (q)}
\left[ \left(  \frac{\partial^2}{\partial p_\alpha \partial p_\nu} - \frac{1}{2}
    \delta_{\alpha\nu} \frac{\partial^2}{\partial p_\beta \partial
      p_\beta} \right) \lb (q+p)_\nu \left[ \phi (q+p) \right] \rb
\e^{{S^*}} \right.\nn\\
&&\left.\qquad - \left(\frac{D-2}{2} + \gamma\right)
  \frac{\partial}{\partial p_\alpha} \left[ 
    \phi (q+p)\right] \e^{{S^*}} \right]
 - \frac{\partial}{\partial p_\alpha} \left( p_\mu p_\nu
    L_{\mu\nu} (p) \right)\nn\\
&=& \int_q K(q) \e^{-{S^*}} \frac{\delta}{\delta \phi (q)} \left[
\left( \lb (q+p)_\nu \frac{\partial^2}{\partial p_\alpha \partial
      p_\nu} - \frac{1}{2} (q+p)_\alpha \frac{\partial^2}{\partial
      p_\beta \partial p_\beta}\rb \left[ \phi (q+p)\right]\right.\right.\nn\\
&&\quad \left.\left. + \left(\frac{D+2}{2}-\gamma\right)
  \frac{\partial}{\partial p_\alpha} 
    \left[ \phi (p+q)\right] \right) \e^{{S^*}} \right]
 -  \frac{\partial}{\partial p_\alpha} \left( p_\mu p_\nu
    L_{\mu\nu} (p) \right)
\end{eqnarray}
In the limit $p \to 0$ the left-hand side vanishes, and we
obtain
\begin{eqnarray}
0 &=& \int_q K(q) \e^{-{S^*}} \frac{\delta}{\delta \phi (q)} \left[
\lb \left(  q_\nu \frac{\partial^2}{\partial q_\alpha \partial q_\nu}
    - \frac{1}{2} q_\alpha \frac{\partial^2}{\partial q_\beta \partial q_\beta}
\right) \left[ \phi (q)\right]\right.\right.\nn\\
&&\left.\left. \qquad+ \left(\frac{D+2}{2}-\gamma\right)
    \frac{\partial}{\partial q_\alpha} 
        \left[ \phi (q)\right] \rb \e^{{S^*}} \right]
\label{eq-conformal}
\end{eqnarray}
This gives the Ward identity for conformal invariance
\begin{eqnarray}
&&\sum_{i=1}^n \left( p_{i\nu} \frac{\partial^2}{\partial
      p_{i\alpha} \partial p_{i\nu}} - \frac{1}{2} p_{i\alpha}
    \frac{\partial^2}{\partial p_{i\beta} \partial p_{i\beta}} + D
    \frac{\partial}{\partial p_{i\alpha}}\right)
\vvev{\phi (p_1) \cdots \phi (p_n)}_{S^*} \nn\\
&&= \left(\frac{D-2}{2} +
    \gamma \right) \sum_{i=1}^n \frac{\partial}{\partial
  p_{i\alpha}} \vvev{\phi (p_1) \cdots \phi (p_n)}_{S^*}
\end{eqnarray}
where the right-hand side is determined by the full scale
dimension of the scalar field at the fixed point.  

If (\ref{O-quadratic}) does not hold, the left-hand side of
(\ref{eq-conformal}) becomes $K_\alpha (0)$, and we obtain
\begin{eqnarray}
&&\sum_{i=1}^n \left( p_{i\nu} \frac{\partial^2}{\partial
      p_{i\alpha} \partial p_{i\nu}} - \frac{1}{2} p_{i\alpha}
    \frac{\partial^2}{\partial p_{i\beta} \partial p_{i\beta}} + D
    \frac{\partial}{\partial p_{i\alpha}}\right)
\vvev{\phi (p_1) \cdots \phi (p_n)}_{S^*} \nn\\
&&= \left(\frac{D-2}{2} +
    \gamma \right) \sum_{i=1}^n \frac{\partial}{\partial
  p_{i\alpha}} \vvev{\phi (p_1) \cdots \phi (p_n)}_{S^*}
- \vvev{K_\alpha (0)\,\phi (p_1) \cdots \phi (p_n)}_{S^*}
\end{eqnarray}
instead.

\section{The Gaussian Fixed Point\label{section-gaussian}}

In this section we would like to give a concrete, though very simple,
example of constructing the symmetric energy-momentum tensor using
(\ref{EM-translation}).

We consider the Wilson action at the gaussian fixed point,
corresponding to the free massless theory with $\gamma = 0$:
\begin{equation}
S_G [\phi] = - \frac{1}{2} \int_p \frac{p^2}{K(p)} \phi (p) \phi (-p)
\end{equation}
We then obtain
\begin{eqnarray}
\left[ \phi (p) \right] &\equiv& \frac{1}{K(p)} \left( \phi (p) +
    \frac{K(p)\left(1-K(p)\right)}{p^2} \frac{\delta S_G}{\delta \phi
      (-p)}\right) = \phi (p)\,,\\
\N (p) &\equiv& - \int_q K(q) \e^{-S_G} \frac{\delta}{\delta \phi (q)}
\left( \phi (p+q) \e^{S_G} \right)\nn\\
&=& \int_{p_1,p_2} \delta (p_1 + p_2 - p)\, \frac{1}{2} \phi
(p_1) \phi (p_2) \, \left( p_1^2 + p_2^2 \right)
\end{eqnarray}
where a field independent constant proportional to $\delta (p)$ has
been ignored for $\N (p)$.  (Composite operators at $S_G$ have been
discussed extensively in Appendix of \cite{Wilson:1973jj}.)
(\ref{EM-translation}) gives
\begin{eqnarray}
p_\mu \Theta_{\mu\nu} (p) &=& \int_q K(q) \e^{- S_G}
\frac{\delta}{\delta \phi (q)} \left( (p+q)_\nu \phi (p+q) \e^{S_G}
\right)\nn\\
&=& - \int_q (p+q)_\nu \phi (p+q) \phi (-q)
\label{EM-translation-G}
\end{eqnarray} 
Now, $\Theta_{\mu\nu} (p)$ is a symmetric tensor with scale dimension
$0$, and it can be written as
\begin{equation}
\Theta_{\mu\nu} (p) = \int_{p_1, p_2} \delta (p_1+p_2 - p)
\frac{1}{2} \phi (p_1) \phi (p_2) \, C_{\mu\nu} (p_1,p_2)
\end{equation}
where
\begin{equation}
C_{\mu\nu} (p_1,p_2) =  A \left(p_{1\mu} p_{1\nu} + p_{2\mu} p_{2 \nu}
    \right)
+ B \left( p_{1\mu} p_{2\nu} + p_{1 \nu} p_{2\mu} \right) +
\delta_{\mu\nu} \left( C \left(p_1^2 + p_2^2\right) + D p_1 p_2 \right)
\end{equation}
Substituting this into (\ref{EM-translation-G}), we can determine $B,
C, D$ in terms of $A$ as
\begin{equation}
C_{\mu\nu} (p_1, p_2) = \delta_{\mu\nu} p_1 p_2 - p_{1\mu} p_{2\nu} -
p_{1\nu} p_{2\mu} + A \lb (p_1+p_2)_\mu (p_1+p_2)_\nu - (p_1+p_2)^2
\delta_{\mu\nu} \rb
\end{equation}
where $A$ is left arbitrary.  Hence, we obtain the energy-momentum tensor
\begin{equation}
\Theta_{\mu\nu} (p) = \delta_{\mu\nu} \left[ \frac{1}{2}
    \frac{1}{i} \partial_\alpha \phi \frac{1}{i} \partial_\alpha \phi
\right] (p) - \left[\frac{1}{i} \partial_\mu \phi
    \frac{1}{i} \partial_\nu \phi \right] (p)
+ A \left(p_\mu p_\nu - p^2 \delta_{\mu\nu}\right) \left[\frac{1}{2}
    \phi^2\right] (p) \label{EM-G}
\end{equation}
where
\begin{eqnarray}
\left[\frac{1}{i} \partial_\mu \phi \frac{1}{i} \partial_\nu
    \phi\right] (p) &\equiv& \int_{p_1, p_2} \delta (p_1+p_2-p) \phi
(p_1) \phi (p_2) \,p_{1\mu} p_{2\nu}\,,\\
\left[\frac{1}{2} \phi^2\right] (p) &\equiv&\int_{p_1, p_2} \delta
(p_1+p_2-p) \,\frac{1}{2} \phi (p_1) \phi (p_2)
\end{eqnarray}
The term proportional to $A$ corresponds to
\begin{equation}
Y_{\mu\alpha,\nu\beta} (p) = A \left(\delta_{\mu\beta} \delta_{\nu\alpha}
    - \delta_{\mu\nu} \delta_{\alpha\beta}\right) \left[\frac{1}{2}
    \phi^2\right] (p) 
\end{equation}
which is a composite operator of scale dimension $-2$.  Since 
the trace is given by
\begin{equation}
\Theta (p) = (D-2) \left[ \frac{1}{2} \frac{1}{i} \partial_\alpha \phi
    \frac{1}{i} \partial_\alpha \phi \right] (p) + A (1-D) p^2  \left[\frac{1}{2}
    \phi^2\right] (p) 
\end{equation}
and the equation-of-motion composite operator by
\begin{equation}
\N (p) = p^2 \left[\frac{1}{2} \phi^2\right] (p) - 2 \left[
    \frac{1}{2} \frac{1}{i} \partial_\alpha \phi 
    \frac{1}{i} \partial_\alpha \phi \right] (p) 
\end{equation}
we obtain
\begin{eqnarray}
\Op (p) &=& \Theta (p) + \frac{D-2}{2} \N (p)\nn\\
&=& \left( A (1-D) + \frac{D-2}{2} \right) p^2 \left[ \frac{1}{2}
    \phi^2 \right] (p)
\end{eqnarray}
which is quadratic in $p$.  Hence, as is well known, the free massless
theory has conformal invariance.  By choosing
\begin{equation}
A = \frac{D-2}{2(D-1)}
\end{equation}
we obtain an improved energy-momentum tensor for which $\Op (p)$
vanishes identically.\cite{Callan:1970ze}

In Appendix \ref{appendix-infinitesimal} we consider an infinitesimal
neighborhood of $S_G$ and construct the energy-momentum tensor there.

\section{Conclusions\label{section-conclusions}}

In this paper we have considered how to construct the energy-momentum
tensor $\Theta_{\mu\nu} (p)$, given a Wilson action which is invariant
under translations and rotations.  To make our task manageable we have
introduced certain restrictions on the kind of Wilson actions we
consider.  We have assumed the continuum description of a Wilson
action which is defined for an arbitrary continuous scalar field in
$D$-dimensional space.  In particular we have assumed that the
ultraviolet cutoff of the theory is provided by a smooth cutoff
function $K (p)$ of squared momentum $p^2$ which is itself rotation
invariant.

Considering how long we have been familiar with the idea of Wilson
actions, it is somewhat surprising that a problem as fundamental as
construction of the energy-momentum tensor for a given Wilson action
has never been considered fully before.  It may be also a little
surprising but reassuring that the naive translation invariance
(\ref{bare-translation-inv}) and rotation invariance
(\ref{bare-rotation-inv}) of a Wilson action give us enough to
construct the energy-momentum tensor $\Theta_{\mu\nu} (p)$ with
expected properties, simply by following the existing formalism.  In
particular, we have derived the Ward identities (\ref{EM-translation})
and (\ref{EM-rotation}), which amount to
\begin{eqnarray}
&&\vvev{p_\mu \Theta_{\mu\nu} (p)\, \phi (p_1) \cdots \phi (p_n)}_S\nn\\
&&\qquad= - \sum_{i=1}^n (p_i+p)_\nu \vvev{\phi (p_1) \cdots \phi (p_i+p)
  \cdots \phi (p_n)}_S\\
&&\vvev{p_\nu \left(\frac{\partial \Theta_{\nu\mu} (p)}{\partial
        p_\alpha} - \frac{\partial \Theta_{\nu\alpha} (p)}{\partial
        p_\mu} \right) \,\phi (p_1) \cdots \phi (p_n)}_S\nn\\
&&\qquad= \sum_{i=1}^n \vvev{\phi (p_1) \cdots \left(
(p+p_i)_\alpha \frac{\partial}{\partial p_\mu} -
(p+p_i)_\mu \frac{\partial}{\partial p_\alpha} \right) \phi (p+p_i)
\cdots \phi (p_n)}_S
\end{eqnarray}
for the correlation functions.  In demonstrating the existence of such
$\Theta_{\mu\nu} (p)$, all we need is the assumption (\ref{locality})
that a local composite operator that vanishes at zero momentum is a
spatial derivative.

We have shown that the symmetry $\Theta_{\mu\nu} (p) = \Theta_{\nu\mu}
(p)$ and the translation invariance (\ref{EM-translation}) determine
$\Theta_{\mu\nu} (p)$ implicitly but almost uniquely.  An additive
ambiguity of the form (\ref{EM-ambiguity}) exists no matter what
formalism we use.

Though we have considered only scalar theories in this paper, it
should be straightforward to generalize our construction of the
energy-momentum tensor to theories with spinor fields.  For gauge
theories, especially YM theories, we may need extra work to
incorporate gauge invariance of the energy-momentum tensor.

\appendix

\section{Equation-of-motion Composite Operators\label{appendix-EOM}}

In this appendix we summarize, for the reader's convenience, the
salient features of the equation-of-motion composite operators, which
were originally called redundant operators in \cite{Wegner:1974} in
the context of the renormalization group.  (The recent review article
\cite{Rosten:2010vm} adopts this original nomenclature.)  More details
can be found in \S 4 of \cite{Igarashi:2009tj}.

Given a composite operator $\Op (p)$, we define the
corresponding equation-of-motion composite operator by
\begin{equation}
\mathcal{E}_{\Op} (p) \equiv - \int_q K(q) \e^{-S[\phi]}
\frac{\delta}{\delta \phi (q)} \left( \Op (q+p) \e^{S [\phi]} \right)
\end{equation}
This has the following modified correlation functions:
\begin{eqnarray}
\vvev{\mathcal{E}_{\Op} (p)\,\phi (p_1) \cdots \phi (p_n)}_S &\equiv&
\prod_{i=1}^n \frac{1}{K (p_i)}\cdot
\vev{\mathcal{E}_{\Op} (p)\, \phi (p_1) \cdots \phi (p_n)}_S\nn\\
&=& \sum_{i=1}^n \vvev{\phi (p_1) \cdots \Op (p_i+p) \cdots \phi
  (p_n)}_S
\end{eqnarray}

For example, if we choose $\Op (p) = \left[\phi (p)\right]$, we obtain
\begin{equation}
\mathcal{E}_{[\phi]} (p) = \N (p) \equiv
- \int_q K(q) \e^{-S[\phi]} \frac{\delta}{\delta \phi (q)} \left(
    \left[ \phi (q+p)\right] \, \e^{S[\phi]} \right)
\end{equation}
satisfying
\begin{equation}
\vvev{\N (p)\,\phi (p_1) \cdots \phi (p_n)}_S
= \sum_{i=1}^n \vvev{\phi (p_1) \cdots \phi (p_i+p) \cdots \phi
  (p_n)}_S
\end{equation}
Especially, $\N \equiv \N (0)$ counts the number of fields:
\begin{equation}
\vvev{\N\, \phi (p_1) \cdots \phi (p_n)}_S = n \vvev{\phi (p_1) \cdots
  \phi (p_n)}_S
\end{equation}

We obtain another example by choosing
\begin{equation}
\Op (p) = p_\mu \frac{\partial}{\partial p_\mu} \left[ \phi (p)
\right]
\end{equation}
We then obtain
\begin{equation}
\vvev{\mathcal{E}_{\Op} (p)\,\phi (p_1) \cdots \phi (p_n)}_S
= \sum_{i=1}^n \vvev{\phi (p_1) \cdots (p + p_i)_{\mu} \frac{\partial
    \phi (p+p_i)}{\partial p_\mu} \cdots \phi (p_n)}_S
\end{equation}
Especially, for $p=0$, we obtain
\begin{equation}
\vvev{\mathcal{E}_{\Op} (0)\,\phi (p_1) \cdots \phi (p_n)}_S
= \sum_{i=1}^n \vvev{\phi (p_1) \cdots p_{i\mu} \frac{\partial
    \phi (p_i)}{\partial p_{i\mu}} \cdots \phi (p_n)}_S
\end{equation}
  
\section{ERG Differential Equations\label{appendix-ERGdiff}}

In \cite{Sonoda:2015bla} it was shown that (\ref{equivalence}) is
equivalent to the ERG differential equation satisfied by $S_t \equiv
R_t S$:
\begin{eqnarray}
\partial_t \e^{S_t [\phi]} &=&
\int_p \left[ \left(\frac{\Delta (p)}{K(p)} + \frac{D+2}{2} - \gamma +
        p_\mu \frac{\partial}{\partial p_\mu} \right) \phi (p) 
    \frac{\delta}{\delta \phi (p)}\right.\nn\\
&& \left.\quad + \frac{1}{p^2}\lb\Delta (p) - 2 \gamma K(p)
        \left(1- K(p)\right)\rb \frac{1}{2} \frac{\delta^2}{\delta
  \phi (p) \delta \phi (-p)} \right] \e^{S_t [\phi]}
\label{ERGdiff}
\end{eqnarray}
where 
\begin{equation}
\Delta (p) \equiv - 2 p^2 \frac{d}{dp^2} K(p)
\label{Delta}
\end{equation}
We wish to rewrite this equation as an operator equation using
equation-of-motion composite operators.

We first rewrite (\ref{equivalence}) as
\begin{equation}
\vvev{\phi (p_1) \cdots \phi (p_n)}_{S_t} = \e^{t \cdot n
  \left(-\frac{D+2}{2} + \gamma \right)} \vvev{\phi (p_1 \e^{-t})
  \cdots \phi (p_n \e^{-t})}_S
\end{equation}
We then differentiate the above with respect to $t$ to obtain
\begin{eqnarray}
\vvev{\partial_t S_t\, \phi (p_1) \cdots \phi (p_n)}_{S_t}
&=& n \left(- \frac{D+2}{2} + \gamma \right) \vvev{\phi (p_1) \cdots
  \phi (p_n)}_{S_t}\nn\\
&& - \sum_{i=1}^n \vvev{\phi (p_1) \cdots p_{i,\mu} \frac{\partial
    \phi (p_i)}{\partial p_{i,\mu}} \cdots \phi (p_n)}_{S_t}
\end{eqnarray}
Hence, using the two examples given in Appendix
\ref{appendix-EOM}, we obtain an operator equation
\begin{equation}
\partial_t S_t [\phi]
= \left(- \frac{D+2}{2} + \gamma\right) \N
+ \int_p K(p) \e^{-S_t [\phi]} \frac{\delta}{\delta \phi (p)} \left(
p_\mu \frac{\partial \left[\phi (p)\right]}{\partial p_\mu} \,\e^{S_t [\phi]}
\right) \label{ERGdiff-rewritten}
\end{equation}
This is equivalent to (\ref{ERGdiff}).  Hence, the Wilson action
changes by an equation-of-motion operator under ERG.

The change of a composite operator under ERG is also an
equation-of-motion operator.  Let $\Op (p)$ be a generic composite
operator.  Its ERG transformation is given by (\ref{Rt-composite}):
\begin{equation}
\vvev{(R_t \Op) (p \e^t) \,\phi (p_1) \cdots \phi (p_n)}_{S_t}
= \e^{t \cdot n \left(- \frac{D+2}{2} + \gamma \right)} \vvev{\Op (p)
  \,\phi (p_1 \e^{-t}) \cdots \phi (p_n \e^{-t})}_S
\end{equation}
Differentiating this with respect to $t$, we obtain
\begin{eqnarray}
&&\vvev{\e^{-S_t} \partial_t \left( (R_t \Op) (p \e^t)\, \e^{S_t}
  \right) \,\phi (p_1) \cdots \phi (p_n)}_{S_t} \nn\\
&&= n \left( - \frac{D+2}{2} +
  \gamma \right) \vvev{(R_t \Op) (p\e^t) \,\phi (p_1) \cdots \phi
(p_n)}_{S_t}\nn\\
&&\quad - \sum_{i=1}^n \vvev{(R_t \Op) (p\e^t) \,\phi (p_1) \cdots p_{i,\mu}
  \frac{\partial \phi (p_i)}{\partial p_{i,\mu}} \cdots \phi
  (p_n)}_{S_t}
\end{eqnarray}
This amounts to the operator equation
\begin{eqnarray}
&&\e^{-S_t} \partial_t \left( (R_t \Op) (p \e^t)\, \e^{S_t} \right)\nn\\
&&= \int_q K(q) \e^{- S_t [\phi]} \frac{\delta}{\delta \phi (q)}
\lb \left(\frac{D+2}{2}-\gamma + q_\mu \frac{\partial}{\partial q_\mu}
\right) \left[ (R_t \Op) (p \e^t)\, \phi (q)\right] \,\e^{S_t [\phi]}
\rb
\label{Rt-operator-equation}
\end{eqnarray}
where $\left[ \Op' \,\phi (p)\right]$ is a composite operator defined by
\begin{equation}
\left[ \Op'\, \phi (p)\right] \equiv \Op'\, \left[ \phi (p) \right] +
\frac{1-K(p)}{p^2} \frac{\delta \Op'}{\delta \phi (-p)} 
\end{equation}
By definition, we find
\begin{equation}
\vvev{ \left[ \Op' \phi (p)\right] \phi (p_1) \cdots \phi (p_n)}_{S_t}
= \vvev{\Op'\, \phi (p) \phi (p_1) \cdots \phi (p_n)}_{S_t}
\end{equation}
The right-hand side of (\ref{Rt-operator-equation}) is an
equation-of-motion composite operator.

At a fixed point $S^*$, a composite operator $\Op$ with scale dimension $-y$
satisfies
\begin{equation}
(R_t \Op) (p \e^t) = \e^{y t} \Op (p \e^t)
\end{equation}
Substituting this into (\ref{Rt-operator-equation}), we obtain
\begin{equation}
\left(y + p_\mu \frac{\partial}{\partial p_\mu}\right) \Op (p)
= \int_q K(q) \e^{- S^*} \frac{\delta}{\delta \phi (q)}
\lb \left(\frac{D+2}{2}-\gamma+q_\mu \frac{\partial}{\partial
      q_\mu}\right) \left[ \Op (p) \phi (q) \right] \e^{S^*} \rb
\label{comp-scale-dim}
\end{equation}

\section{$\Theta_{\mu\nu} (p)$ in the Infinitesimal Neighborhood of $S_G$\label{appendix-infinitesimal}}

We consider an infinitesimal neighborhood of the gaussian fixed point
$S_G$ by adding an arbitrary potential term with infinitesimal coefficients:
\begin{equation}
S \lb g\rb [\phi] = S_G [\phi] - \sum_{n=1}^\infty g_n
\left[\frac{1}{(2n)!} \phi^{2n} \right] (0)
\end{equation}
where
\begin{eqnarray}
\left[ \frac{1}{(2n)!} \phi^{2n} \right] (p)
&\equiv& \sum_{k=0}^{n-1} \frac{C^k}{k!} \int_{p_1,\cdots, p_{2(n-k)}}
\delta (p_1 + \cdots + p_{2(n-k)}-p)\nn\\
&& \quad \times \frac{1}{(2(n-k))!}\, \phi (p_1) \cdots \phi (p_{2(n-k)})
\end{eqnarray}
is a composite operator of scale dimension $- y_n\equiv n(D-2)-D$ defined at
$S_G$.  The constant $C$ is defined by
\begin{equation}
C \equiv - \frac{1}{2} \int_q \frac{K(q)}{q^2}
\end{equation}
so that $\left[ \phi^{2n}\right]$ is normal ordered.  We find
\begin{equation}
\vvev{\left[ \phi^{2n} \right] (p \e^t) \,\phi (p_1 \e^t) \cdots \phi
  (p_k \e^t)}_{S_G} = \e^{t \left( - y_n - n \frac{D+2}{2} \right)}
\vvev{\left[ \phi^{2n} \right] (p)\, \phi (p_1) \cdots \phi
  (p_k)}_{S_G}
\end{equation}

The symmetric energy-momentum tensor can be constructed from
(\ref{EM-translation}).  We only give results here.  To first order in
$g$'s, the energy-momentum tensor is given by
\begin{eqnarray}
\Theta_{\mu\nu} (p) &=& \delta_{\mu\nu} \left[ \frac{1}{2}
    \frac{1}{i} \partial_\alpha \phi \frac{1}{i} \partial_\alpha \phi
\right] (p) - \left[\frac{1}{i} \partial_\mu \phi
    \frac{1}{i} \partial_\nu \phi \right] (p)\nn\\
&& + A \left(p_\mu p_\nu - p^2 \delta_{\mu\nu}\right) \left[\frac{1}{2}
    \phi^2\right] (p) - \delta_{\mu\nu} \sum_{n=1}^\infty g_n
\left[\frac{1}{(2n)!} \phi^{2n} \right] (p)
\end{eqnarray}
where the composite operators are corrected to first order in $g$'s as
\begin{eqnarray}
    \left[\frac{1}{i} \partial_\mu \phi \frac{1}{i} \partial_\nu
        \phi\right] (p) &=& \int_{p_1,p_2} \delta (p_1+p_2-p) \phi (p_1)
    \phi (p_2) p_{1\mu} p_{2\nu}  - \tilde{J}_{\mu\nu} (p)
    \sum_{n=2}^\infty  \tilde{g}_n  \int_{p_1,\cdots,p_{2(n-1)}}\nn\\
    && \times \delta (p_1 + \cdots + p_{2(n-1)}-p)
    \frac{1}{\left(2(n-1)\right)!}  \phi (p_1) \cdots \phi
    (p_{2(n-1)})\\ 
    \left[\frac{1}{2} \phi^2\right] (p) &=& \int_{p_1,p_2} \delta
    (p_1+p_2-p) \frac{1}{2} \phi (p_1) \phi (p_2)
    - \tilde{I} (p) \sum_{n=2}^\infty \tilde{g}_n  \int_{p_1,\cdots,p_{2(n-1)}}\nn\\
    &&
    \times \delta 
    (p_1 + \cdots + p_{2(n-1)}-p) \frac{1}{\left(2(n-1)\right)!} 
    \phi (p_1) \cdots \phi (p_{2(n-1)})
\end{eqnarray}
The parameter $\tilde{g}_n$ is defined by
\begin{equation}
\tilde{g}_n \equiv \sum_{k=0}^\infty g_{n+k} \frac{C^k}{k!}
\end{equation}
The functions $\tilde{I} (p)$ and $\tilde{J}_{\mu\nu} (p)$ are well
defined for $2 < D < 4$, and given by
\begin{eqnarray}
\tilde{I} (p) &\equiv& \frac{1}{2} \int_q \frac{1-K(q)}{q^2} \frac{1 -
  K(q+p)}{(q+p)^2}\\
\tilde{J}_{\mu\nu} (p) &\equiv& \int_0^\infty dt\, \e^{t(D-2)} \left(
2 \int_q \frac{\Delta (q)}{q^2} \frac{1 - K(q + p \e^{-t})}{(q+p
  \e^{-t})^2} (q + p \e^{-t})_\mu (-q)_\nu \right.\nn\\
&&\left.+ \frac{2}{D} \delta_{\mu\nu} \int_q \frac{\Delta
  (q)\left(1-K(q)\right)}{q^2} \right)
 + \frac{2}{D(D-2)} \delta_{\mu\nu} \int_q \frac{\Delta
  (q)\left(1-K(q)\right)}{q^2}
\end{eqnarray}
where $\Delta (p)$ is defined by (\ref{Delta}).

The trace is given, again to first order, by
\begin{equation}
\Theta (p) = - \frac{D-2}{2} \N (p) - \sum_{n=1}^\infty y_n g_n \left[
    \frac{1}{(2n)!} \phi^{2n}\right] (p)
\end{equation}
if $A = (D-2)/(2(D-1))$ is chosen.  Note that $y_n g_n$ is the beta
function of $g_n$ under ERG.

\begin{acknowledgments}
    I would like to thank Prof.~Nicol\'{a}s Wschebor of Universit\'e
    Pierre et Marie Curie and Universidad de la Rep\'ublica, Uruguay
    for an inspiring talk at ERG2014 (Lefkada, Greece), which
    motivated me to work on this subject.  This work was partially
    supported by the JSPS grant-in-aid \# 25400258.
\end{acknowledgments}

\bibliography{paper-v2}

\end{document}